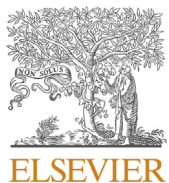

Research article

# Skills in computational thinking of engineering students of the first school year

Concepción Varela, Carolina Rebollar, Olatz García, Eugenio Bravo, Javier Bilbao[*]

*Applied Mathematics Department, University of the Basque Country, Pl. Ing. Torres Quevedo, 1, 48013, Bilbao, Spain*

ABSTRACT

In this world of the digital era, in which we are living, one of the fundamental competences that students must acquire is the competence in Computational Thinking (CT). Although there is no general consensus on a formal definition, there is a general understanding of it as a set of skills and attitudes necessary for the resolution, with or without a computer, of problems that may arise in any area of life. Measuring and evaluating which of the CT skills students have acquired is fundamental, and for this purpose, previously validated measuring instruments must be used. In this study, a previously validated instrument is applied to know if the new students in the Engineering Degrees of the University of the Basque Country have the following skills in CT: Critical Thinking, Algorithmic Thinking, Problem Solving, Cooperativity and Creativity.

## 1. Introduction

In order to be successful in the changing world in which we live, we need to be adaptable, sensitive to changes, to be able to solve problems and develop software and hardware, or, where appropriate, use and create technology. Regardless of the area in which they are going to work, students need to prepare themselves for the future with skills in Communication, Collaboration, Creativity, Critical Thinking, Computing. That is, they need to acquire the competence of Computational Thinking (CT) [dataset] (ISTE et al., 2011).

Contrary to what we could believe, the CT is not only applicable in the field of computer science or robotics [dataset] (Liang et al., 2013; García-Peñalvo and Mendes, 2018; Rojas-López and García-Peñalvo, 2018; Rojas-López and García-Peñalvo, 2019). Learning to think in a computational way, beginning with learning to use abstraction as a basic tool of reasoning, brings with it a series of educational benefits that reinforce intellectual skills and improve the learning of any subject.

In the nineties, the concept that was used was "Digital Literacy", and the first approach that exists about it is given to [dataset] Paul (Gilster, 1997), who explained it as "the ability to understand and use information in multiple formats from a wide variety of sources when it is presented via computers".

Jeannette [dataset] Wing, 2006 used the term "computational thinking" to articulate a vision that everyone, not just those who major in computer science, can benefit from thinking like a computer scientist, but afterwards gave the following definition of CT: "Computational Thinking is the thought processes involved in formulating problems and their solutions so that the solutions are represented in a form that can be effectively carried out by an information-processing agent" [dataset] (Wing, 2010).

When Critical Thinking is combined with the power of computing, a solid base is established to make decisions and to innovate solutions that can improve the resolution of problems of any kind that may arise. It should be noted that the power of computing is unlimited and that it will lead us to increase our capacity to solve problems to a level that nowadays we can not even imagine [dataset] (ISTE et al., 2011).

We are not only talking about great problems solved by researchers, innovation or technology development, we are also talking about thinking in such a way, when we solve problems, that we can rely on technology, on digital tools, if necessary or as we wish. Think, for example, about the kitchen robot, which can help us improve the quality of life or, of course, in many other cases that robots help us to be more efficient in the work we develop. Let's think about the future work that we will develop taking into account the technological advances that we will have available.

We must admit the need to prepare people to acquire the CT competence, and the most effective way to do this is by integrating it into the curricula of the compulsory education. In this sense, some initiatives are already underway in various parts of the world in order to integrate CT into teaching, as reflected in the Computhink report [dataset] (Ferrari et al., 2018), which gives an overview of those countries that, or either they already have this competence in their curriculum, or are planning to include it. The Autonomous Community of the Basque Country is also working in this line: its department of Education has the digital competence included, within the transversal or generic core competences, in the HEZIBERRI 2020 plan [dataset] (Education Department, 2014).

In this work, we have studied the CT competence of the students who have started their studies in the engineering degrees of the University of the Basque Country, using the scale designed and validated statistically by [dataset] Korkmaz et al. (2017).

## 2. Theoretical framework

### 2.1. Computational thinking

Although it has been a while since Wing gave a first definition of CT, it is continuously being revised without having reached a consensus yet. Thus, for [dataset] (ISTE and CSTA, 2011), computational thinking is a problem-solving process that includes (but is not limited to) the following characteristics:

- Formulating problems in a way that it is possible to use a computer and other tools in their resolution.
- Organizing data logically and analyzing them.
- Representing data through abstractions.
- Automating solutions.
- Identifying, analyzing and implementing possible solutions in order to obtain the most effective combination.
- Generalizing and transferring this process to a wide variety of solutions.

Computational thinking can also be understood as a general term that encompasses cognitive skills involved in computational tasks [dataset] (Doleck et al., 2017).

The design of tasks to implement CT is based on those skills that you want to measure. Those that are usually included in the literature are the following: abstraction, data analysis, debugging, algorithmic thinking, cooperativity, creativity, critical thinking, problem solving, recursive thinking and heuristic thinking [dataset] (Barr and Stephenson, 2011; Brennan and Resnick, 2012; Román et al., 2015; Korkmaz et al., 2017).

However, the implementation of CT would not be complete if it was not evaluated. That is, you need to know which the competences acquired by the students are. Here, there is a big gap because in order to carry out a correct evaluation, the measurement instrument or questionnaire to be used must be previously validated [dataset] (Grover and Pea, 2013).

Within the literature that addresses this problem, the following measurement instruments have been found to evaluate CT in primary school courses and pre-university courses: "*Fairy Assessment in Alice*", a tool designed to measure algorithmic thinking, abstraction and modeling [dataset] (Werner et al., 2012); "*Computational Thinking Framework (CTF)*", reference that is used both to validate materials and to evaluate students [dataset] (Gouws et al., 2013). In Taiwan, a large-scale competition has been developed to evaluate the CT in the pre-university courses studying computer and student fluency, and for this purpose, 15 tasks have been chosen from the 'International Bebras Contest' [dataset] (Bebras, 2018). In Spain, [dataset] Román et al. (2015) have designed a CT test based mainly on the concepts of computing and using the syntax of computer languages [dataset] (Román et al., 2017)).

At university level, the University of Kentucky has designed an instrument to analyze the relationship between critical thinking and CT for a subject of Informatics. In a first application, they used an instrument designed by themselves, not validated, and that did not give good results with respect to critical thinking. Therefore, in a second application, the questions related to critical thinking were obtained from The California Critical Thinking Disposition Inventory (CCTDI) and the California Critical Thinking Skills Test (CCTST) [dataset] (Facione and Facione, 1992), validated instrument, and, to measure CT, they used the one they had designed in the first stage with questions related to computational concepts [dataset] (Walden et al., 2013). In South Africa, to know the academic performance of first-year university students [dataset] (Gouws et al., 2013), they have carried out a study with questions taken from '*Computer Olympiad Talent Search*', which aims to show both the ability to program using different languages such as C ++, Java or Pascal, and teamwork [dataset] (Román et al., 2015). In Turkey, they have designed a test that is valid and reliable to measure the following CT skills: Creativity, Algorithmic Thinking, Critical Thinking, Problem Solving and Cooperativity [dataset] (Korkmaz et al., 2017), which are discussed below.

### 2.2. Creativity

"Creativity is intelligence having fun", Albert Einstein.

Creativity is defined as the "ability to create" or the "capacity for creation"". In addition, the definition of creating refers to "producing something from nothing", "establishing, founding, introducing something for the first time; giving birth to it or giving it life, in a figurative sense". That is, when we talk about creativity we refer to new things, ideas or concepts. Moreover, we associate creativity with originality, innovation and imagination. However, as indicated by [dataset] Bruner (1962), together with the idea of novelty, two other concepts associated with creativity must not be ignored, such as the relevance of the created being or idea and its effectiveness. That is, no one would say that a person is creative for the simple fact of having created something new, if it is recognized that this creation has no value, material or aesthetic or of any kind, nor if we cannot give it any use.

In 1950, Guilford [dataset] establishes a definition of Creative Thinking based on differentiating two models of thought: convergent and divergent. He associated the first one with a traditional way of reasoning, vertical, that adapts to what is established and follows a logical criterion. The second one is what he associated, in fact, with Creative Thinking. It follows a horizontal reasoning, works in several planes at once and it questions the established. The characteristics of this divergent thinking would be:

- Fluency: ability to offer several alternatives to a problem.
- Flexibility: ability to develop the same issue in different ways.
- Originality: ability to create new ideas.
- Redefinition: ability to think without being influenced by the known.
- Penetration: ability to discover new factors in the analyzed problem.
- Preparation: ability to generate detailed information on the subject matter.

The idea of promoting Creative Thinking among the students should not be identified with an education lacking in norms, imprecise, in which any behavior is admitted and in which the acquisition of other basic skills is left out. At the other end, it should not be identified with an elitist or pretentious education, which is expected to obtain small geniuses, although the impulse of Creativity in the classroom will contribute something in that way [dataset] (Cropley, 1995).

The new challenges of the future, many still unknown, require new approaches and solutions, hence the importance of Creative Thinking. Creative Thinking can be considered as the necessary skill to understand a new problem, giving it the necessary focus, so that its resolution is addressed in the best way. In that sense, Creative Thinking is not an isolated skill, but has a close relationship with the skills of Problem Solving and Algorithmic Thinking [dataset] (Korkmaz et al., 2017). We could say that, in order to develop an optimal resolution algorithm, a prior analysis of the proposed problem is essential, and in that analysis is where Creative Thinking can play an important role.

## 2.3. Algorithmic thinking

Algorithmic Thinking is the ability to understand, execute, evaluate and create algorithms [dataset] (Brown, 2015). On the other hand, an algorithm is a set of instructions that, executed step by step, in a certain order, represents a model to solve a task. Often, algorithmic thinking is also called Logical-Mathematical Thinking. According to Gardner, in his studies on multiple intelligences, Logical-Mathematical Thinking is associated with the skills of recognizing patterns, reasoning deductively and thinking with logic. According to [dataset] Linn et al. (2010), computational thinking is closely related to logical-mathematical thinking, but it is not exactly the same. Although in both cases great importance is given to the capacity of abstraction, in the case of Mathematical Thinking that abstraction is associated to the used structures (equations), while in the case of Computational Thinking that abstraction extends to the methodology.

## 2.4. Critical thinking

Our quality of life depends on our thinking because it is the ability that allows us to make decisions and solve problems. Thought is part of human nature, however, by itself, our thinking is arbitrary, distorted, biased, uninformed or prejudiced. Poor quality thinking leads to a poor quality of life. To achieve a better quality of life, one must exercise thought in a systematic way. Pestalozzi said 200 years ago that thinking directs man towards knowledge. You can see, hear and read what you want and as much as you want; but you will never know anything about it, except for what you have thought about; about what, because you have thought about it, you have made it property of your own mind.

According to [dataset] (Elder and Paul, 2007), the most useful definition for assessing critical thinking skills is the following: "Critical thinking is the process of analyzing and assessing thinking with a view to improving it. Critical thinking presupposes knowledge of the most basic structures in thinking (the elements of thought) and the most basic intellectual standards for thinking (universal intellectual standards). The key to the creative side of critical thinking (the actual improvement of thought) is in restructuring thinking as a result of analyzing and effectively assessing it".

A critical and cultivated thinker:

- formulates problems and vital questions, with clarity and precision.
- accumulates and evaluates relevant information and uses abstract ideas to interpret that information effectively.
- reaches conclusions and solutions, testing them with relevant criteria and standards.
- thinks with an open mind within alternative systems of thought; recognizes and evaluates, as necessary, the assumptions, implications and practical consequences and
- when devising solutions to complex problems, this thinker communicates them effectively.

In summary, critical thinking is self-directed, self-disciplined, self-regulated and self-corrected. It means to submit to rigorous standards of excellence and conscious mastery of its use. It implies effective communication and problem solving skills and a commitment to overcome the egocentricity and natural centrism of the human being.

- All reasoning has a purpose.
- All reasoning is an attempt to solve a problem, settle a question or explain something.
- All reasoning is based on assumptions.
- All reasoning is done from one perspective.
- -All reasoning is based on data, information and evidence.
- All reasoning is expressed through concepts and ideas that simultaneously shape it.
- All reasoning contains inferences or interpretations by which we arrive at conclusions and that give meaning to data.
- All reasoning leads somewhere, has implications and consequences.

## 2.5. Problem solving

Information and Communication Technologies (ICT) have revolutionized our social ways. They have made possible a new global and dynamic environment (the digital environment), in which we socially interact, exercise our rights as citizens, do science and new business models that make other traditional ones obsolete emerge [dataset] (Echeverría, 1999). Citizens and professionals need to acquire the necessary skills to get on in this environment. They must be able to face complex problems, which are not even clearly defined during their training period. And they should know how to take advantage of the opportunities that the environment provides them.

"Problem-solving competence is defined as the capacity to engage in cognitive processing to understand and resolve problem situations where a method of solution is not immediately obvious. It includes the willingness to engage with such situations in order to achieve one's potential as a constructive and reflective citizen" [dataset] (OECD, 2014).

According to the study "Soft skills 4 talent 2016" [dataset] (Human Age Institute, 2016), the most related skill to Talent is Problem Solving. Problem Solving is the most valued social competence for 69% of those responsible for human resources surveyed in that study, followed by Orientation to Objectives (58%) and Collaboration (57%). Those people with high capacity for solving problems are able to act proactively, without wasting time, and finding the most appropriate solutions for each case, always thinking about the repercussions they may have in the long term.

In a similar way, according to the research work "Solving Chemistry Problems and Cognitive Structures" conducted at the University of Keele (U.K.) [dataset] (Kempa, 1986), two main ways of conceiving problem solving and understanding its function are distinguished. The first one involves productive processes of problem solving, in which the person produces by discovery a combination of previously learned rules that can be applied to obtain a solution to a novel situation or problem. The second one involves reproductive processes in which knowledge is simply remembered or applied in non-novel situations or problems. Computational Thinking addresses the resolution of problems as a productive process, through pedagogical proposals focused on authentic problems adjusted to the reality of the students. This allows them to face different situations without repeating predetermined solutions. Based on the interrelation of previous knowledge, students develop novel solutions and adjusted to the characteristics of the situation they are facing [dataset] (Artecona et al., 2017).

## 2.6. Cooperativity

Computational thinking has usually been considered as an individual competence. However, at present, real problems are becoming increasingly complex problems in which a single person finds great difficulty in finding the solution. In addition, in Engineering and other areas of knowledge, it is usual to work in groups and having to coordinate and cooperate with other people from the same or different department or company.

Therefore, we need people who are able to work as a team. This cooperative work allows obtaining as a result a greater capacity than the sum of the individual capacities of each one of the members of the group, when it comes to solving problems.

According to [dataset] Missiroli et al. (2017) "cooperative thinking is the ability to describe, recognize, decompose problems and computationally solve them in teams in a socially sustainable way". This definition joins together the concepts of CT and cooperative learning.

In the literature, the concepts of cooperation and collaboration have been found indistinctly. The definitions given are as follows [dataset] (Roschelle and Teasley, 1995):

- "Cooperative work is accomplished by the division of labor among participants, as an activity where each person is responsible for a portion of the problem solving".
- "Collaboration involves the mutual engagement of participants in a coordinate effort to solve the problem together".

This concept of collaboration does not rule out the possibility of distributing tasks among group participants. Moreover, the most prepared component to perform a certain task usually proposes him/herself when performing this task.

The definition of collaboration seems the most appropriate to define collaborative or cooperative learning.

This method of learning is not only interesting to improve learning success, but it also facilitates the information exchange among students and improves their communication skills within a group.

In this way, the collaboration between individuals with different abilities when solving problems has become a fundamental aspect of work in our days, so it must also become one of the most important skills that students should acquire as throughout their formative process.

## 3. Method

### 3.1. Sample

The study involved 1138 first-year students of the UPV/EHU of the different campuses that make up the university: Bizkaia, Gipuzkoa and Araba. Gathering of the information has been carried out during the first academic week of the course 2018/19. 59.8% of the sample belongs to the Bizkaia Campus, 19.9% to the Gipuzkoa Campus and 20.3% to the Araba Campus. The participants mostly study in Spanish, 61.1%, compared to 38.9% who study in Basque. The distribution by gender, and the degree to which the participants have registered, is shown in Table 1.

The lack of response in a questionnaire is a problem to be addressed prior to carrying out any statistical analysis. In this sample, the percentage of partial non-response is high and represents 13% of the total sample, so removing those individuals who have not answered any of the questions in the questionnaire is not the most appropriate, and therefore, it is necessary to apply an imputation method. In this case, the "linear trend at point" method, provided by the statistical software SPSS, has been imputed. This method replaces missing values with the linear trend for that point and the existing series is regressed on an index variable scaled 1 to n, where missing values are replaced with their predicted values [dataset] (IBM Corp, 2016).

### 3.2. Description of the questionnaire

When a questionnaire is used as an instrument for the statistical exploitation of data, it must be well designed according to the standard quality criteria. For this reason, and for analytical convenience, in this study we have used the questionnaire designed by [dataset] Korkmaz et al. (2017) to measure CT skills. For the generation of this questionnaire, and according to the ability of CT to be measured, we have used different instruments with acceptable psychometric qualities. Briefly, the questionnaire is designed to measure "Creativity" (from [dataset] Whetton and Cameron (2002) and adapted by [dataset] Aksoy (2004)), "Problem solving" skill ([dataset] Heppner and Petersen (1982)), the ability of "Cooperation" ([dataset] Korkmaz (2012)), "Critical thinking" ([dataset] (California Academic Press LLC, 2018)), and the "Algorithmic Thinking" competence ([dataset] Yesil and Korkmaz (2010)). The questionnaire was revised by experts and, after performing the Exploratory Factor Analysis (EFA) and the corresponding Confirmatory Factor Analysis (CFA), 29 questions or items were finally selected to validate the Computational Thinking [dataset] (Korkmaz et al., 2017) of the students distributed as follows: 6 for measuring algorithmic thinking, 4 for cooperativity, 5 for critical thinking, 6 to measure problem solving and, 8 for creativity. Each item has 6 possible ratings according to a Likert scale: "(1) strongly disagree", "(2) disagree", "(3) neither agree nor disagree", "(4) agree", "(5) strongly agree", and the score (6) has been added to collect information from those who do not want or do not know what to answer.

### 3.3. Data analysis

Factorial Analysis (FA) is a data reduction technique that is used to find homogeneous groups of variables that explain, in a simplified way, the information that is in a large set of variables. That is to say, what FA intended is to simplify the information given by a correlation matrix in order to be able to be easily interpreted [dataset] (Fernández Aráuz, 2015a,b; Pérez Gil et al., 2000 Lloret Segura et al., 2014). For this reason, it is a widely used statistical technique for the validation of questionnaires with satisfaction scales.

From a conceptual point of view, there are two types of FA: the exploratory (EFA) and the confirmatory (CFA). The use of one type or the other one will depend on the objective of the research to be carried out. The EFA tries to find factors from the interpretive task and attributes a posteriori a meaning to the factors. The CFA, however, implies explicitly specifying a model about the underlying factors and it is subjected to confirmation with the observed data [dataset] (López-Roldán and Fachelli, 2015a).

As the questionnaire used in this study has already been validated previously with the relevant analyzes, and, following the criteria of [dataset] Henson and Roberts (2006) (the questionnaire is not newly created and the factorial structure is known in other samples), we chose to initially use the CFA to check whether the structure of the 5 factors or latent variables confirmed by [dataset] Korkmaz et al. (2017) is also revealed in the used sample. For the extraction of the factors, the Maximum Likelihood Estimation is used.

When the CFA does not produce the desired results, it would proceed as follows [dataset] (Brown, 2006):

- The sample will be divided into two random halves, m1 and m2 respectively.
- In the sample m1, EFA will be done.
- In the sample m2, CFA will be implemented to verify the results obtained in the previous step.

**Table 1**
Distribution according to the degree of study and gender.

| | | Gender | | | Total |
|---|---|---|---|---|---|
| | | Female | Male | Non-binary | |
| Degree | IndTech, Org, Env | 124 | 202 | 0 | 326 |
| | Teleco | 25 | 62 | 0 | 87 |
| | Civil | 20 | 22 | 0 | 42 |
| | Inform | 9 | 67 | 0 | 76 |
| | Elec, Mec, El&Auto | 20 | 126 | 0 | 146 |
| | Gipuz | 41 | 182 | 3 | 226 |
| | Automo | 2 | 37 | 0 | 39 |
| | Infor Gaste | 10 | 54 | 1 | 65 |
| | Mec Gaste | 14 | 50 | 0 | 64 |
| | Chem Gaste | 3 | 29 | 1 | 33 |
| | Electro Gaste | 9 | 21 | 0 | 30 |
| Total | | 277 | 852 | 5 | 1134 |

IndTech, Org, Env = Industrial Technology Engineering, Industrial Organization Engineering, Environmental Engineering, at Bizkaia Campus (BC); Teleco = Telecommunications Engineering (BC); Civil = Civil Engineering (BC); Inform = Computer Science (BC); Elec, Mec, El&Auto = Electrical Engineering, Mechanical Engineering, Electronic & Control Engineering (BC); Gipuz = degrees at Gipuzkoa Campus (Civil Engineering, Electrical Engineering, Mechanical Engineering, Electronic & Control Engineering); Automo = Automotive Engineering at Araba Campus (AC); Infor Gaste = Computer Science (AC); Mec Gaste = Mechanical Engineering (AC); Chem Gaste = Chemical Engineering (AC); Electro Gaste = Electronic & Control Engineering (AC).

The realization of the EFA has been done with the principal components method and using the Varimax criterion for analytic rotation [dataset] (Kaiser, 1958), which involves the orthogonal rotation of the latent variables to make obtained factors more interpretable [dataset] (López-Roldán and Fachelli, 2015b). We should mention that an important difference between oblique and orthogonal rotations is that they can create factors that are correlated or uncorrelated with each other: rotations that allow for correlation are called oblique rotations; rotations that assume the factors are not correlated are called orthogonal rotations. In order to choose the appropriate number of factors, the criterion of the eigenvalue greater than one is followed together with the analysis of the sedimentation graph.

The choice of the items that define each factor is based on the weight it has in the factors, considering that an appropriate value would be around 0.4. As additional information, the Index of Fit for Factor Scales (IFFS) has been used [dataset] (Morales Vallejo, 2011; Fleming, 1985).

In order to determine the viability of the CFA, the following statistics are used [dataset] (Brown, 2006):

1. Kaiser-Meyer-Olkin (KMO), which measures the suitability of the data for the realization of the CFA considering that the results of the model will be: excellent if $0.9 < KMO <1$, good if $0.8 < KMO <0.9$ and, acceptable if $0.7 < KMO <0.8$.
2. Barlett's test of sphericity, which contrasts the hypothesis that the correlation matrix is the identity matrix. If the level of significance is $p < 0.05$, the null hypothesis is rejected, accepting the relationship between the variables and the validity of the CFA.

The information provided by the CFA is necessary to know the structure of the factors, but it is not enough. Therefore, together with the confirmatory analysis, Structural Equation Modeling (SEM) is used, which provides more indicators to verify the goodness or correct fit of the proposed model. Next, the used indicators are shown as well as the intervals in which they should be in order to consider that the model has an acceptable adjustment: $\chi^2/d < 5$; $0.06 < RMSEA < 0.08$; $0 \leq S-RMR \leq 0.08$; $0.90 \leq NNFI \leq 0.96$; $0.90 \leq CFI < 0.96$; $0.90 \leq GFI \leq 0.96$; $0.90 \leq AGFI \leq 0.96$ and $0.90 \leq IFI \leq 0.96$ [dataset] (Kline, 2011; Medrano and Muñoz-Navarro, 2017; Hooper et al., 2008; Fernández Aráuz, 2015a,b; Ruiz et al., 2010).

The reliability of the questionnaire was carried out by studying the following analysis:

a) Analysis of internal consistency (how closely related a set of items are as a group), calculating the Cronbach's alpha [dataset] (Cronbach, 1951), based on the average inter-correlation among the items. The values of this coefficient oscillate between 0 and 1, considering it acceptable if it is greater than 0.7. The coefficient has also been analyzed by eliminating each of the items that make up the questionnaire, to check if by eliminating any item, the coefficient will increase.
b) Analysis of the discriminating capacity of the items, analyzing their homogeneity index. The threshold of this index is 0.2 [dataset] (Lacave et al., 2015).

The statistical treatment of the data has been done with the SPSS software, version 24.0, and the structural equation modelling with the AMOS module of that software.

## 4. Results

### 4.1. Findings regarding the validity of the scale of CT skills

On the total of the sample and with the 29 starting items of the questionnaire, the CFA was conducted with five factors [dataset] (Korkmaz et al., 2017). The KMO and Bartlett statistics are shown in Table 2.

The values of the diagonal of the anti-image matrix contain the measure of sample adequacy and they are values between 0.713 and 0.911. All this implies the appropriateness of the realization of the CFA. However, the matrix of factor loads in the 5 factors (Table 3) shows very few values that are, in absolute value, greater than 0.5. In addition, as we can see in the table, regarding to the first, fourth and fifth factors, there is no clear set of items that can define them. Only factor 2 would be defined with the items related to the "Cooperativity" ability, and factor 3 with those related to "Problem solving". These results suggest that there are some items that, in this sample, could be highly correlated; and, therefore, we cannot obtain a clear factorial structure.

Table 4 shows the Pearson correlation coefficients for the items "Algorithmic Thinking" and "Creativity" with a significance of 0.01. Similar results are seen between "Algorithmic Thinking" and "Critical Thinking". That is, for the sample that has been analyzed these items are giving redundant information, which prevents from seeing a clear underlying factorial structure. Therefore, the items related to "Creativity" and "Critical Thinking" are removed.

As we have not found confirmation of the 5 initial factors, and having removed from the analysis the questions related to "Creativity" and "Critical Thinking", we went on to explore the structure of the sample with the questions related to: "Algorithmic Thinking", "Problem Solving" and "Cooperativity". The new scale consists of 16 items from the initial 29.

**Table 2**
Tests of KMO and bartlett.

| Kaiser-Meyer-Olkin measure of sampling adequacy | | ,836 |
|---|---|---|
| Barlett's test of sphericity | Approx. Chi-Square | 6778,453 |
| | df | 406 |
| | Sig. | ,000 |

**Table 3**
Factorial matrix with five factors.

| | Factor | | | | |
|---|---|---|---|---|---|
| | 1 | 2 | 3 | 4 | 5 |
| AT1 | .352 | .207 | .126 | .254 | .027 |
| AT2 | .415 | .215 | .132 | .248 | .012 |
| AT3 | .305 | .143 | .197 | .182 | .072 |
| AT4 | .410 | .251 | .035 | .256 | .058 |
| AT5 | .337 | .119 | .193 | .255 | .075 |
| AT6 | .300 | .112 | .170 | .253 | .167 |
| COOP1 | .458 | **-.524** | -.031 | .014 | -.079 |
| COOP2 | .413 | **-.637** | -.032 | .009 | .026 |
| COOP3 | **.514** | **-.609** | -.069 | .070 | -.032 |
| COOP4 | .439 | **-.303** | -.061 | -.033 | .032 |
| CT1 | .406 | .218 | .168 | .187 | -.062 |
| CT2 | .406 | .209 | .142 | .197 | -.039 |
| CT3 | .443 | .126 | .051 | -.008 | -.045 |
| CT4 | .441 | .158 | .079 | .010 | .052 |
| CT5 | .260 | .071 | .172 | .007 | .111 |
| PS1 | -.142 | -.330 | **.464** | -.080 | -.061 |
| PS2 | -.195 | -.270 | **.527** | -.088 | -.141 |
| PS3 | -.180 | -.285 | **.520** | -.013 | .018 |
| PS4 | -.258 | -.208 | **.550** | -.019 | .031 |
| PS5 | -.267 | .020 | **.566** | .106 | -.016 |
| PS6 | -.246 | .221 | **.404** | .064 | -.066 |
| CREA1 | .262 | .027 | .141 | -.210 | .348 |
| CREA2 | .268 | .066 | .166 | -.328 | .443 |
| CREA3 | **.528** | .277 | .107 | -.310 | -.171 |
| CREA4 | **.551** | .304 | .108 | -.262 | -.220 |
| CREA5 | .443 | .233 | .160 | -.113 | -.091 |
| CREA6 | .334 | .086 | .107 | -.158 | .072 |
| CREA7 | .321 | .122 | .087 | -.018 | -.002 |
| CREA8 | .383 | .138 | .093 | -.037 | -.039 |

Extraction method: maximum likelihood. 5 extracted factors. 9 necessary iterations.

Table 4
Correlations between algorithmic thinking (AT) and creativity (CREA).

|       |                  | AT1     | AT2     | AT3     | AT4     | AT5     | AT6     |
|-------|------------------|---------|---------|---------|---------|---------|---------|
| CREA1 | Correlation      | .059*   | .137**  | .053    | .111**  | .086**  | .107**  |
|       | Sig. (bilateral) | .048    | .000    | .075    | .000    | .004    | .000    |
| CREA2 | Correlation      | .081**  | .055    | .107**  | .075*   | .058*   | .096**  |
|       | Sig. (bilateral) | .006    | .062    | .000    | .012    | .050    | .001    |
| CREA3 | Correlation      | .156**  | .225**  | .150**  | .213**  | .141**  | .105**  |
|       | Sig. (bilateral) | .000    | .000    | .000    | .000    | .000    | .000    |
| CREA4 | Correlation      | .204**  | .227**  | .192**  | .238**  | .171**  | .139**  |
|       | Sig. (bilateral) | .000    | .000    | .000    | .000    | .000    | .000    |
| CREA5 | Correlation      | .208**  | .212**  | .162**  | .245**  | .205**  | .131**  |
|       | Sig. (bilateral) | .000    | .000    | .000    | .000    | .000    | .000    |
| CREA6 | Correlation      | .110**  | .130**  | .131**  | .083**  | .110**  | .101**  |
|       | Sig. (bilateral) | .000    | .000    | .000    | .005    | .000    | .001    |
| CREA7 | Correlation      | .158**  | .127**  | .164**  | .148**  | .138**  | .093**  |
|       | Sig. (bilateral) | .000    | .000    | .000    | .000    | .000    | .002    |
| CREA8 | Correlation      | .137**  | .196**  | .117**  | .143**  | .134**  | .143**  |
|       | Sig. (bilateral) | .000    | .000    | .000    | .000    | .000    | .002    |

** Correlation is significant at level 0.01 (bilateral).
* Correlation is significant at level 0.05 (bilateral).

### 4.2. Exploratory factor analysis

In this phase, the total sample is divided into two subsamples in a random manner, and an EFA is performed in the first sample with 578 participants in order to detect possible latent variables with the items AT1, AT2, AT3, AT4, AT5, AT6, COOP1, COOP2, COOP3, COOP4, PS1, PS2, PS3, PS4, PS5 and PS6. The KMO statistic on the scale is 0.746 and the values of the Barlett's test: $\chi^2 = 1849.055$, df = 120 (p = 0.000). These suggest that there are latent variables or constructs and, therefore, the exploratory factor analysis makes sense.

The principal component analysis shows that the communality of the items is between 0.333 and 0.648 without making rotation (communality is the variance in observed variables accounted for by common factors). On the contrary, these values are between 0.556 and 0.828 after performing the Varimax rotation (there is an item with a deviation of 0.432 in its corresponding factor, but even so it has been decided to keep it in the analysis), being the total variance explained by the rotated factors of 46.1%. The analysis of the eigenvalue greater than one shows that there are three factors, those related to: "Cooperativity" with 4 items, "Algorithmic Thinking" with 6 items and "Problem Solving" also with 6 items. The calculation of the IFFS (Index of Fit of Factor Scales) with values in the three factors above 0.90 verifies that the chosen items define correctly each of the factors.

### 4.3. Confirmatory Factor Analysis

To verify the previous results, a CFA is carried out in the second subsample consisting of 559 participants. The factor analysis is done with three factors and the estimates with the maximum likelihood method. The KMO statistic in the scale is of 0.779 and the values of the Barlett's test $\chi^2 = 2064.570$; df = 120 (p = 0.000). The total variance explained by the three factors is 48.822%. Fig. 1 shows the sedimentation graph, where we can notice the sharp decrease between the first three eigenvalues, and how from the fourth eigenvalue this decrease is more moderate.

As a result of the followed process, Table 5 shows the values of the scale in the different factors. The first factor explains 18.679% of the total variance, the items that contribute to the creation of this construct are COOP1, COOP2, COOP3 and COOP4, with weights ranging from 0.567 to 0.787. The second factor explains 16.608% of the total variance with the items PS1, PS2, PS3, PS4, PS5 and PS6, with weights from 0.56 to 0.647 (except for the one related to the item PS6, with a small weight). And, finally, the third factor explains 13.535% of the variance and it is made up of the items AT1, AT2, AT3, AT4, AT5 and AT6, with weights ranging from 0.460 to 0.571.

The Structural Equations model that arises and that collects the information provided by the CFA is represented in Fig. 2.

When analyzing the goodness of the fit, the following indicators are obtained: $\chi^2/d$ = 2.9; RMSEA = 0.06; NNFI = 0.88; CFI = 0.90; GFI = 0.94; AGFI = 0.92; CFI = 0.90 and IFI = 0.90. Therefore, it follows that the fit model is acceptable, and, consequently, the 3 chosen factors are confirmed. The values of the estimated parameters of the model obtained with the maximum likelihood technique are shown in Table 6.

The correlations among the items and their corresponding factors have also been calculated. These values can be seen in Table 7, confirming that these are significant at a level of 0.01. Therefore, it can be said that each item serves both the purpose of the factor and the questionnaire.

### 4.4. Reliability analysis of the questionnaire

The evaluation of the reliability of the questionnaire was carried out using the Cronbach's alpha coefficient. This value is 0.635 for the questionnaire made up of the 16 questions related to Algorithmic Thinking, Cooperativity and Problem Solving. The index does not improve when deleting variables from the questionnaire. Obtaining the value of the coefficient for each one of the obtained factors, it is verified that the questionnaire measures well the abilities of Cooperativity and Problem Solving, with coefficients of 0.80 and 0.74 respectively, whereas in the case of the skill Algorithmic Thinking, the coefficient is 0.65. With regard to the index of homogeneity (or discrimination) of the items, they generally have a good level, values between 0.30 and 0.39, except for the item PS6 (related to Problem Solving) that has a homogeneity index of 0.05.

## 5. Discussion

In this study, we have applied a scale to know the CT skills (Problem Solving, Algorithmic Thinking, Cooperativity, Critical Thinking and Creativity) that the Engineering students of the University of the Basque Country have. This scale was previously validated by [dataset] Korkmaz et al. (2017). The scale is a Likert scale of 5 values with 29 items and has been carried out simultaneously in the three Historical Territories of the Autonomous Community of the Basque Country. The Confirmatory Factor Analysis, with five factors, performed with the whole sample, has not shown this clear factorial structure, due to the existing correlation between the variables corresponding to Creativity, Critical Thinking and Algorithmic Thinking. This correlation makes that there is a factor in which the variables that contribute to its formation are indistinctly those related to the three aforementioned skills. On the other hand, the questions that appear on the scale related to Creativity are directed to a group of people more heterogeneous than the one that has participated in this study, so it does not seem strange that this ability does not appear clearly in the participants of this study. That is, in the study conducted by Korkmaz et al., 684 students from the sample used in the exploratory factorial analysis (72% of respondents) belong to degrees in Education, Sociology and Psychology. And 580 students from the sample used in the confirmatory factorial analysis belong to ten different grades, related to both Sciences and Humanities. However, in our case, 100% of the grades and students belong to the area of Engineering. Therefore, and without modifying the questionnaire, we wanted to know if the engineering students have the three skills that have been considered basic for this kind of technical studies: Problem Solving, Algorithmic Thinking and Cooperativity.

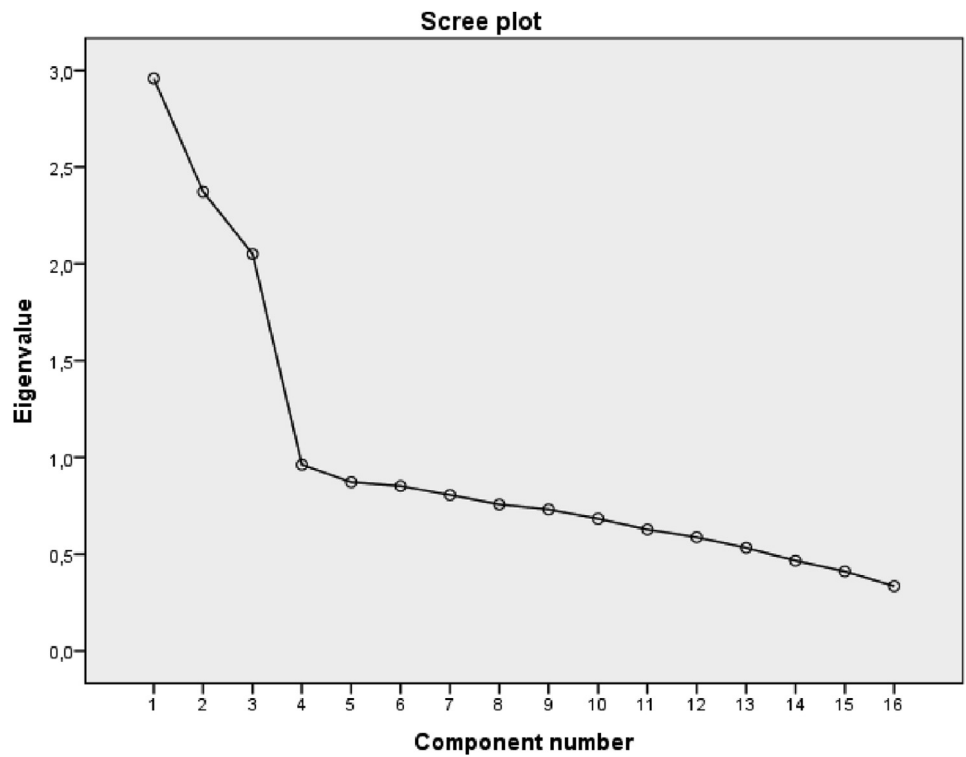

**Fig. 1.** Screen plot graphic.

**Table 5**
Factor analysis of the scale as per factors.

| Items | | F1 | F2 | F3 |
|---|---|---|---|---|
| Cooperativity | In the cooperative learning, I think that I attain/will attain more successful results because I am working in a group. | 0.783 | | |
| | I like experiencing cooperative learning together with my group friends. | 0.779 | | |
| | I like solving problems related to group project together with my friends in cooperative learning | 0.678 | | |
| | More ideas occur in cooperative learning | 0.556 | | |
| Problem Solving | I cannot produce so many options while thinking of the possible solution ways regarding a problem | | 0.647 | |
| | I cannot apply the solution ways I plan respectively and gradually | | 0.645 | |
| | I have problems in the demonstration of the solution of a problem in my mind. | | 0.625 | |
| | I have problems in the issue of where and how I should use the variables such as X and Y in the solution of a problem | | 0.621 | |
| | I cannot develop my own ideas in the environment of cooperative learning | | 0.596 | |
| | It tires me to try to learn something together with my group friends in cooperative learning | | 0.340 | |
| Algorithmic Thinking | I cannot develop my own ideas in the environment of cooperative learning | | | 0.571 |
| | It tires me to try to learn something together with my group friends in cooperative learning | | | 0.546 |
| | I can digitize a mathematical problem expressed verbally. | | | 0.472 |
| | I cannot produce so many options while thinking of the possible solution ways regarding a problem | | | 0.466 |
| | I have problems in the issue of where and how I should use the variables such as X and Y in the solution of a problem | | | 0.460 |
| | I cannot apply the solution ways I plan respectively and gradually | | | 0.460 |
| Eigenvalue | | 2.989 | 2.657 | 2.166 |
| Explained variance | | 18.679 | 16.608 | 13.535 |
| IFFS | | 0.90 | 0.97 | 0.94 |

The total sample of 1138 participants was divided into two random subsamples, to perform an Exploratory Factor Analysis followed by a Confirmatory one. The first one has shown a factorial structure composed of three factors, and the corresponding statistics confirm the relevance of the analysis. The Confirmatory Factor Analysis corroborates these results, as well as the goodness of fit indices of the structural equations model. That is, the proposed model is verified by the data.

Item-factor correlations have been calculated to identify at which level each of the variables on the scale measures the factor to which it belongs. According to the obtained values, it has been found that each variable serves the proposed purpose. The internal consistency of the scale has been studied with Cronbach's alpha and the homogeneity index, giving acceptable results.

The three factors that have been obtained can be explained briefly as follows:

Algorithmic Thinking is the ability to understand, execute, evaluate

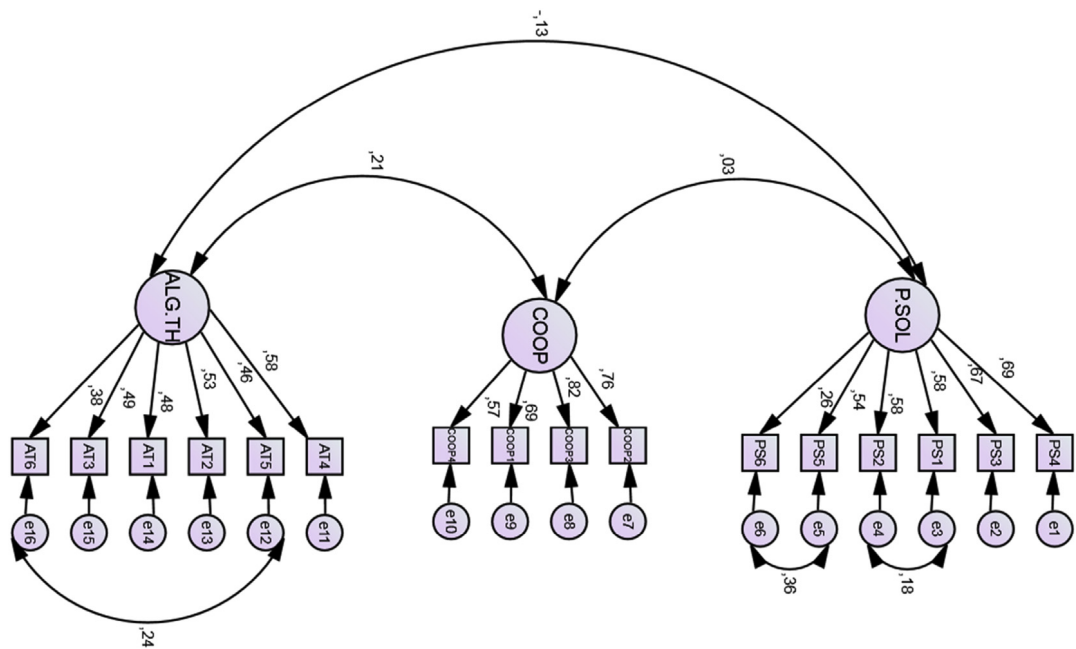

**Fig. 2.** Confirmatory factor analysis diagram of the scale.

Table 6
Standardized regression weights.

| Item | | Estimate | Item | | Estimate |
| --- | --- | --- | --- | --- | --- |
| COOP3 | <— | .821 | AT6 | <— | .379 |
| COOP1 | <— | .685 | AT4 | <— | .577 |
| COOP4 | <— | .571 | PS3 | <— | .666 |
| COOP2 | <— | .756 | PS1 | <— | .582 |
| AT5 | <— | .459 | PS2 | <— | .583 |
| AT2 | <— | .532 | PS5 | <— | .544 |
| AT1 | <— | .478 | PS6 | <— | .279 |
| AT3 | <— | .491 | PS4 | <— | .688 |

Table 7
Item-factor scores correlation analysis.

| F1 Cooperativity | | F2 Problem solving | | F3 Algorithmic thinking | |
| --- | --- | --- | --- | --- | --- |
| I | r | I | r | I | r |
| COOP1 | .744** | PS1 | .705** | AT1 | .563** |
| COOP2 | .859** | PS2 | .700** | AT2 | .570** |
| COOP3 | .855** | PS3 | .728** | AT3 | .557** |
| COOP4 | .610** | PS4 | .730** | AT4 | .691** |
| | | PS5 | .672** | AT5 | .660** |
| | | PS6 | .384** | AT6 | .556** |

N = 559; ** = correlation is significant at level 0.01.

and create algorithms [dataset] (Brown, 2015). Problem Solving is defined as the ability to participate in a cognitive process to understand and solve problems where there is no immediately obvious method of solution [dataset] (OECD, 2014). Cooperative thinking is the ability to describe, recognize, decompose problems and computationally solve them in teams in a socially sustainable way [dataset] (Missiroli et al., 2017).

## 6. Conclusion

It has been analyzed if the engineering students who register in the first year at the UPV/EHU have acquired CT skills. For this purpose, a scale on CT has been used, which has been previously statistically validated. The results suggest that two of the CT abilities, Critical Thinking and Creativity, are not measured in this sample with this scale. The group in which the study was conducted is a homogeneous group with respect to the studies prior to entering the university, and they have a strong background in STEM subjects, so it is not surprising that the two mentioned skills appear mixed within those that have been statistically validated in the sample.

It cannot be said that engineering students accessing the university for the first time lack the abilities of creativity or critical thinking. The conclusion is that another scale would be needed, valid and reliable from the statistical point of view, to measure these skills together with problem solving, algorithmic thinking and cooperativity in engineering students.

## Declarations

### Author contribution statement

J. Bilbao, C. Varela, C. Rebollar, O. Garcia, E. Bravo: Conceived and designed the experiments; Performed the experiments; Analyzed and interpreted the data; Contributed reagents, materials, analysis tools or data; Wrote the paper.

### Funding statement

This research did not receive any specific grant from funding agencies in the public, commercial, or not-for-profit sectors.

### Competing interest statement

The authors declare no conflict of interest.

### Additional information

No additional information is available for this paper.